\documentclass[showpacs,preprintnumbers,amsmath,amssymb]{revtex4}
\usepackage{graphicx}
\usepackage{dcolumn}
\makeatletter
\input{epsf}

\newcommand{\bd}{\begin{document}}
\newcommand{\ed}{\end{document}}
\newcommand{\bc}{\begin{center}}
\newcommand{\ec}{\end{center}}
\newcommand{\bfr}{\begin{flushright}}
\newcommand{\efr}{\end{flushright}}
\newcommand{\vs}{\vspace}
\newcommand{\hs}{\hspace}
\newcommand{\beq}{\begin{equation}}
\newcommand{\eeq}{\end{equation}}
\newcommand{\lb}{\linebreak}
\newcommand{\mb}{\makebox}
\newcommand{\fb}{\framebox}
\newcommand{\mc}{\multicolumn}
\newcommand{\ben}{\begin{enumerate}}
\newcommand{\een}{\end{enumerate}}
\newcommand{\bit}{\begin{itemize}}
\newcommand{\eit}{\end{itemize}}
\newcommand{\un}{\underline}
\newcommand{\lefq}{\lefteqn}
\newcommand{\ba}{\begin{array}}
\newcommand{\ea}{\end{array}}
\newcommand{\beqa}{\begin{eqnarray}}
\newcommand{\eeqa}{\end{eqnarray}}
\newcommand{\beqas}{\begin{eqnarray*}}
\newcommand{\eeqas}{\end{eqnarray*}}
\newcommand{\bfg}{\begin{figure}}
\newcommand{\efg}{\end{figure}}
\newcommand{\bds}{\begin{displaymath}}
\newcommand{\eds}{\end{displaymath}}
\newcommand{\btb}{\begin{tabbing}}
\newcommand{\etb}{\end{tabbing}}
\newcommand{\para}{\parallel}
\newcommand{\pad}{\partial}
\newcommand{\nn}{\nonumber}
\newcommand{\la}{\leftarrow}
\newcommand{\ra}{\rightarrow}
\newcommand{\lgla}{\longleftarrow}
\newcommand{\lgra}{\longrightarrow}
\newcommand{\La}{\Leftarrow}
\newcommand{\Ra}{\Rightarrow}
\newcommand{\Lra}{\Leftrightarrow}
\newcommand{\Lgla}{\Longleftarrow}
\newcommand{\Lgra}{\Longrightarrow}
\newcommand{\bm}{\boldmath}
\newcommand{\lan}{\langle}
\newcommand{\ran}{\rangle}
\renewcommand{\a}{\alpha}
\renewcommand{\b}{\beta}
\newcommand{\g}{\gamma}
\newcommand{\G}{\Gamma}
\renewcommand{\d}{\delta}
\newcommand{\eps}{\epsilon}
\newcommand{\s}{\sigma}
\newcommand{\lam}{\lambda}
\newcommand{\D}{\Delta}
\newcommand{\vare}{\varepsilon}
\newcommand{\pr}{\prime}
\newcommand{\ro}{\rho}
\newcommand{\nab}{\nabla}
\newcommand{\m}{\mu}
\newcommand{\n}{\nu}
\newcommand{\Sg}{\Sigma}
\newcommand{\p}{\pi}
\newcommand{\R}{I\!\!R}
\newcommand{\om}{\omega}
\newcommand{\Om}{\Omega}
\newcommand{\ze}{\zeta}
\newcommand{\vart}{\vartheta}
\newcommand{\tri}{\triangle}
\newcommand{\f}{\frac}
\newcommand{\iny}{\infty}
\newcommand{\pro}{\propto}
\begin{document}
%\preprint{APS/123-QED}
\title{Quantum Hall Skyrmions in the framework of
O(4) Non-linear Sigma Model}

\author{B. Basu}
\email{banasri@isical.ac.in}
\author{S. Dhar}
 \email{sarmishtha_r@isical.ac.in}
\author{P. Bandyopadhyay}
 \email{pratul@isical.ac.in}
\affiliation{Physics and Applied Mathematics Unit\\
 Indian Statistical Institute\\
 Kolkata-700108 }
%\date{\today}

\begin{abstract}
A new framework for quantum Hall skyrmions in $O(4)$ nonlinear
sigma model is studied here. The size and energy of the skyrmions
are determined incorporating the quartic stability term in the
Lagrangian. Moreover, the introduction of a $\theta$-term
determines the spin and statistics of these skyrmions.
\end{abstract}

\maketitle

\section{Introduction}

Skyrmions are generally described by nonlinear sigma model. A
smooth texturing of the spins can be described by an effective
nonlinear sigma model where the spin is described as a unit vector
${\bf{n}}({\bf{r}})$. This spin vector ${\bf{n}}$ lies on a sphere
$S^{2}$ where we may consider a magnetic monopole located at the
centre. Skyrmions are characterized by the topological charge
$Z=\int d {\bf{r}} q({\bf{r}})$ where
$q={\bf{n}}.(\partial_{x}{\bf{n}}\times\partial_{y}{\bf{n}})/4\pi$
is the Pontryagin density. Z is the winding number of the mapping
${\bf{n}}({\bf{r}})$ from the compactified space $S^2$ to the
target space of the sigma model $(S^2)$ and is given by the
homotopy $\pi_{2}(S^2)=Z$. The topological density is proportional
to the deviation of the electron density from its uniform value
$\rho_{0}$.

In analogy to the Ginzburg-Landau description of
superconductivity, the Chern-Simons theory of quantum Hall systems
was derived by Zhang, Hansson and Kivelson \cite{5} which was
subsequently extended by Lee and Kane \cite{6} to describe the
spin unpolarized quantum Hall liquid. In a quantum Hall fluid, the
ground state has been found to be a fully spin polarized quantum
ferromagnet. By noticing that the dynamics of quantum Hall system
with a spin polarized ground state will follow that of a quantum
ferromagnet and that the skyrmion is a charged object of the
system, Sondhi et.al. \cite{s1,s2} proposed a phenomenological
action which is valid for the long wave length and small frequency
limit. In this scheme, the competition between the Zeeman and
Coulomb terms sets the size and energy of the skyrmions and
modifies the detailed form of their profiles.  In case the Zeeman
energy and Coulomb energy are taken to be vanishing, analytic
expressions are available for the skyrmions and their energy is
independent of their size. This scale invariance is broken by the
Zeeman and Coulomb terms in the Lagrangian.

To study low energy excitations for partially polarized states a
nonlinear $\sigma$-model has been developed  \cite{7} considering
a two component quantum Hall system within a Landau-Ginzberg
theory with two Chern-Simons gauge fields. It has been found that
these excitations have finite energy due to the presence of the
Chern-Simons gauge field and closely resemble the skyrmions in the
usual nonlinear $\sigma$-model. Skyrmions in arbitrary polarized
quantum Hall states are studied \cite{ssm} employing a doublet
model taking into account two fields carrying the spin index in
Chern-Simons term with a matrix valued coupling strength.

Skyrmion excitations in quantum Hall systems at $\nu=1$, using
finite size calculations \cite{xh}, is studied in spherical
geometry where electrons reside on the surface of a sphere with a
monopole at the centre. For a monopole of total magnetic flux
$N_\phi =2\mu$ the lowest Landau level degeneracy is $2\mu+1$. All
single particle states in lowest Landau level have fixed angular
momentum value $\mu$. At filling factor $\nu=1$ corresponding to
the electron number $N=2\mu+1$ the ground state is spin polarized
independent of the strength of the Zeeman splitting. When an
additional flux is added or removed from the system, the ground
state is a skyrmion (antiskyrmion) with topological charge
$Q=+1(-1)$. The energy gap is determined by the interplay of the
Zeeman energy and electron-electron interaction.

In this note, we shall study the static properties of skyrmions in
spherical geometry considering a $O(4)$ nonlinear $\sigma$-model
in $3+1$ dimensional manifold. In some earlier papers
\cite{rev11,bb} we have analyzed the hierarchy of quantum Hall
states in spherical geometry from the view point of chiral anomaly
and Berry phase. In this framework it is shown that the Magnus
force acting on vortices and skyrmions in the quantum Hall system
is generated by the background field associated with the chirality
of the system \cite{bdm}. The various polarization states of
quantum Hall fluid is also studied and their low energy
excitations are examined using a homotopic analysis \cite{bdp}.
The $O(4)$ nonlinear $\sigma$-model helps us to have two
independent subalgebras each isomorphic to a $SU(2)$ algebra so
that the left and right group can be taken to be associated with
two mutually opposite orientations of the magnetization vector
which resides on the $2D$ surface of the sphere. The homotopic
analysis then suggests that for fully polarized states the low
energy excitations are skyrmions. It is here shown that the
addition of the quartic stability term introduced by Skyrme, known
as the Skyrme term in the literature, will determine the size of
the quantum Hall skyrmions in this framework.  This also helps us
to determine the spin and statistics of the skyrmion through the
introduction of the topological $\theta$-term in the Lagrangian.

In sec.2 we recapitulate the study of skyrmion excitations in
quantum Hall fluid when the system is described by means of the
Zhang-Hansson-Kivelson model and modified by Lee and Kane to take
into account the effect of spin. In sec.3 we shall describe
quantum Hall skyrmions in spherical geometry considering a $O(4)$
nonlinear $\sigma$-model taking into account the Skyrme term and
$\theta$-term. In sec.4 we shall determine the size and energy of
the skyrmion in this formalism.

\section{Quantum Hall Skyrmions in Planar Geometry}

We here recapitulate the works on quantum Hall skyrmions on the
basis of the bosonic theory of an electron which is  viewed as a
composite object of a boson and a flux tube carrying an odd
multiple of flux quanta $\Phi_{0}=h/e$ attached via a Chern-Simons
term. We begin with the Landau-Ginzburg theory of the Hall effect
introduced by Zhang, Hansson and Kivelson \cite{5} and modified by
Lee and Kane \cite{6} to incorporate the effect of the spin. We
consider the Lagrangian \cite{st}

\begin{eqnarray}
L=&&i\phi^\dag[\partial_{0}-i(a_{0}+eA_{0})]\phi\nonumber\\
&&-\frac{1}{2m^*}|[\partial_{i}-i(a_{i}+eA_{i})]\phi|^2\nonumber\\
&&-\frac{\lambda}{2}(|\phi|^2-\rho_{0})^2
+\frac{1}{4\theta}\epsilon^{\mu\nu\sigma}a_{\mu}\partial_{\nu}a_{\sigma}
\end{eqnarray}

Here $\phi=(\phi_{1},\phi_{2})$ is a two component complex scalar
field. $\theta$ is the statistics  parameter which takes the
values $\theta=(2n+1)\pi$ so that the boson field $\phi$
represents a fermion, $A_{\mu}$ the external electromagnetic field
and $m^{*}$ is the effective mass of the electron. To determine
the size and energy of the skyrmions Zeeman and Coulomb
interactions should be included. However, to study the topological
features we can ignore them for the moment. We consider a solution
with uniform density $\rho=(|\phi_1|^2+|\phi_2|^2)=\rho_0$ at
filling fraction $\nu=\frac{1}{2n+1}$. In order to separate  the
amplitude and spin degrees of freedom we introduce the CP$^1$
field $\bf{z}$ and we write

\begin{equation}
\phi=\sqrt{\rho}\bf{z}
\end{equation}

The Lagrangian (1) now takes the form

\begin{eqnarray}
L=&&i\rho{\bf{z}}^{\dag}[\partial_{0}-i(a_0+eA_0)]{\bf{z}}\nonumber\\
&&-\frac{\rho}{2m^*}|[\partial_i-i(a_i+eA_i)]{\bf{z}}|^2\nonumber\\
&&-\frac{\lambda}{2}(\rho-\rho_0)^2+\frac{1}{4\theta}\epsilon^{\mu\nu\sigma}a_{\mu}\partial_{\nu}a_{\sigma}
\end{eqnarray}
Here $\bf{z}$ is a two component complex field with the constraint
${\bf{z}}^{\dag}{\bf{z}}=|z_1|^2+|z_2|^2=1$.

One can now use the identity

\begin{equation}
\frac{\rho}{2m^*}|[\partial_i-i(a_i+eA_i)] {\bf{z}
}|^2=\frac{\rho}{8m^*} (\bigtriangledown {\bf{n}}
)^2+\frac{m^*}{2\rho} \bf{J} ^2
\end{equation}
where $ {\bf{J}} =(J^1,J^2)$ with

\begin{equation}
J^i=\frac{\rho}{m^*i}[ {\bf{z}} ^{\dag}\partial_i {\bf{z}}
-i(a_i+eA_i)]
\end{equation}

The local spin direction is given by

\begin{equation}
n^a={\bf{z}} ^{\dag}\sigma^a {\bf{z}}
\end{equation}
and $\nabla{\bf{n}}=(\partial_1{\bf{n}},\partial_2{\bf{n}})$. The
term $\frac{\rho}{8m*}(\nabla{\bf{n}})^2$ corresponds to the
static nonlinear sigma model.

Now introducing Hubbard-Stratanovich transformation ${\bf{J}}$
with the relation $J_0=\rho$ so that the current 3-vector
$(J^0,J^1,J^2)$ is promoted to the status of an independent
dynamical variable, the Lagrangian can be written as

\begin{equation}
\begin{array}{lcl}
L &= &\displaystyle{i\rho[{\bf{z}}^{\dag}\partial_0{\bf{z}}
-i(a_0+eA_0)]}\\
&&\displaystyle{+i[{\bf{z}}^{\dag}\partial_i{\bf{z}}
-i(a_i+eA_i)]J^i+\frac{m^*}{2\rho}{\bf{J}}^2}\\
&&\displaystyle{-\frac{\rho}{8m*}(\nabla{\bf{n}})^2
-\frac{\lambda}{2}(\rho-\rho_0)^2+\frac{1}{4\theta}\epsilon^{\mu\nu\lambda}a_{\mu}\partial_{\nu}a_{\lambda}}\\
\end{array}
\end{equation}

Integrating over the $U(1)$ phase degree of freedom in ${\bf z}$
we arrive at the current conservation law
$\partial_{\alpha}J_{\alpha}=0$ and we have current 3-vector
$(\rho=J^0,J^1,J^2)$ as the curl of a three dimensional vector
field. One can set $J^{\mu}_{[0]}=(\rho_0,0,0)$ equal to
$\epsilon^{\mu\nu\sigma}\partial_{\nu}{\cal{A}}^{[0]}_{\sigma}$
and

\begin{equation}
J^{\mu}-J^{\mu}_{[0]}=\epsilon^{\mu\nu\sigma}\partial_{\nu}\mathcal{A}_{\sigma}
\end{equation}

Integrating out the Chern-Simons field $a_{\mu}$ and using the
relation $2\theta\rho_0=eB_z$, we can finally arrive at the
relation

\begin{equation}
\begin{array}{lcl}
L &=&\displaystyle{2\pi
{\cal{J}}^{\mu}_S({\cal{A}}_{\mu}+{\cal{A}}^{[0]}_{\mu})
-\theta {J^{\mu}} {\cal{A}}_{\mu}}\\
&&\displaystyle{+\frac{m^{\ast}}{2\rho} {\bf{J}}^2
-\frac{\rho}{8m^{\ast}}(\nabla{\bf{n}})^2
-\frac{\lambda}{2}(J^0-J^0_{[0]})^2}\\
\end{array}
\end{equation}
where

\begin{equation}
{\mathcal{J}}^{\mu}_S= {\frac{1}{2{\pi} i}}
\epsilon^{\mu\nu\sigma}\partial_{\mu}\bar{z}_{\sigma}\partial_{\nu}z_{\sigma}
\end{equation}
is the skyrmion number current.

 Defining the field strength term~
$\cal{F}_{\mu\nu}=\partial_{\mu}\cal{A}_{\nu}-\partial_{\nu}\cal{A}_{\mu}$~
to be the dual of the electron number current and adjusting the
unit of length and time such that  $c=\sqrt{\lambda
\rho_0/m^{*}}$, the velocity of density waves in the absence of
magnetic field becomes unity, we can write

\begin{equation}
\begin{array}{lcl}
L &=&\displaystyle{2\pi
{\cal{J}}^{\mu}_S({\cal{A}}_{\mu}+{\cal{A}}^{[0]}_{\mu})-\frac{1}{2}\theta\epsilon^{\mu\nu\sigma}{\cal{A}}_{\mu}{\cal{F}}_{\nu\sigma}}\\
&&\displaystyle{-\frac{\lambda}{4}{\cal{F}}_{\mu\nu}{\cal{F}}^{\mu\nu}-\frac{1}{8\lambda}(\nabla{\bf{n}})^2}\\
\end{array}
\label{11}
\end{equation}

It is noted that the skyrmion number current acts as a source for
a topologically massive gauge field ${\mathcal{A}}_{\mu}$
\cite{st} and also sees the background field
${\mathcal{A}}^{[0]}_{\mu}$.

Incorporating Zeeman and Coulomb terms Sondhi et.al. \cite{s1}
considered the following modified form of the Lagrangian (1).

\begin{equation}
\begin{array}{lcl}
L({\bf{r}})
&=&\displaystyle{\phi^{\dag}({\bf{r}})[\partial_0-i(a_0+eA_0)]\phi({\bf{r}})]}\\
&&\displaystyle{-\frac{1}{2m^*}|[\partial_i-i(a_i+eA_i)]\phi({\bf r})|^2}\\
&&\displaystyle{-\frac{1}{2}\int d^2 r^{\prime}
V({\bf{r}}-{\bf{r}}^{\prime})[|\phi({\bf{r}})|^2-\rho_0][|\phi({\bf{r}}^{\prime})|^2-\rho_0]}\\
&&\displaystyle{-\frac{e^2}{4\theta}\epsilon^{\mu\nu\sigma}a_{\mu}({\bf{r}})\partial_{\nu}a_{\sigma}({\bf{r}})
-\frac{1}{2}g \mu_B
B\phi^{\dag}({\bf{r}})\sigma^z\phi({\bf{r}})}\\
\end{array}
\end{equation}
Here  $V=e^2/\epsilon r$ is the interparticle Coulomb potential
and $\mu_B$ is the Bohr magneton.

Noting that $\phi_{\alpha}=\sqrt{\rho}z_{\alpha}$ where $\sum
{z_{\a}}^{\dag} z_{\a}=1$ and using the mapping
$n^a={\bf{z}}^{\dag}\sigma^a{\bf{z}}$, we can replace the CP$^1$
field by an $O(3)$ sigma model field. Indeed, by observing that at
the ground state the dynamics of the system is that of a
ferromagnet with a long range interaction arising from the Coulomb
interaction between the underlying electrons, we take into account
the necessary terms and consider the Lagrangian \cite{st}
\begin{equation}
  \begin{array}{lcl}
L_{eff} &=&\displaystyle{\frac{1}{2}\rho^0
{\bf{A}}({\bf{n}}).\partial_{t}{\bf{n}}-\frac{1}{2}\rho^{s}(\nabla{\bf{n}})^2}\\
&&\displaystyle{+\frac{1}{2}g \rho_0 \mu_B
{\bf{n}}.{\bf{B}}-\frac{e^2}{2\epsilon}\int
d^2 r^{\prime}\frac{q({\bf{r}})q({\bf{r}}^{\prime})}{|{\bf{r}}-{\bf{r}}^{\prime}|}}\\
\end{array}
\label{13}
\end{equation}
Here $\bf{A}$ is the vector potential of a unit magnetic monopole,
$\rho^s$ is the spin stiffness
($\rho^s={e^2}/{16\sqrt{2\pi}\epsilon l}$ for $\nu=1$), $\epsilon$
is the dielectric constant and $l$ is the magnetic length.
$q({\bf{r}})={\bf{n}}.(\partial_x {\bf{n}}\times
\partial_y {\bf{n}})/4\pi$ is the topological density.
This topological density is proportional to the deviation of the
electron density $\rho$ from its uniform value;
$q=\nu(\rho-\rho_0)$. The topological charge is $Z=\int
d{\bf{r}}q({\bf{r}})$. For $\nu=1$, skyrmions with topological
charge $Z$ carry electric charge $-Z|e|$.

The Zeeman and Coulomb terms break the scale invariance. Their
profiles now depend on the dimensionless ratio
 $\tilde{g}=(g\mu_B B)/(e^2/\epsilon l^2)$ of the Zeeman energy to Coulomb energy.
For $Z=1$ skyrmions the scale invariant solution yields a Zeeman
energy that diverges logarithmically with system size for any
$\lambda$. This can be fixed by matching the scale invariant
solution onto the exact asymptotic solution in the outer region.
This suggest for $Z=1$ the size \cite{s3}

\begin{equation}
{\lambda}=0.558~ l(\tilde{g}|\ln\tilde{g}|)^{-1/3}
\end{equation}
 and energy
\begin{equation}
E=\frac{e^2}{\epsilon
l}~[\sqrt{\frac{\pi}{32}}+0.622~(\tilde{g}|\ln\tilde{g}|)^{1/3}]
\end{equation}

Indeed at large $g$ the quasiparticles are single particle-like
and may carry charge and spin $S_z=1/2$ and have size $l$. At
small $g$, they still carry charge $\pm e$ but diverge in size and
have nontrivial spin with a divergent $z$-component of spin $S_z$
(the number of reversed spin) as well as divergent total spin.

\section{Quantum Hall Skyrmions in $3+1$ dimensional manifold}

We want to study the low energy topological excitations of the
quantum Hall fluid within the framework of $O(4)$ nonlinear
$\sigma$-model  in $3+1$ dimensional manifold. In this geometry,
electrons are placed on the surface of a sphere under the
influence of an uniform radial magnetic field. The magnetic field
is produced by a magnetic monopole of strength $\mu$ placed at the
centre of the sphere. As we know, the angular momentum in the
field of a magnetic monopole is given by

\begin{equation}
{\bf{J}}={\bf{r}}\times {\bf{p}}-\mu{\bf{\hat{r}}}
\end{equation}
where $\mu$  can take the values $0,\pm 1/2,\pm 1,\pm
3/2,........$. Evidently, eigenvalues of ${\bf J}$ can take the
values $|\mu|$, $|\mu|+1$, $|\mu|+2$,....... In this geometry the
quantum Hall  states can be depicted by a two-component spinor
$\left(\begin{array}{c}
  u \\
  v
\end{array}\right)$
such that $u~=~\cos{\theta/2}~e^{i\phi/2}$ and
$v~=~\sin{\theta/2}~e^{-i\phi/2}$ \cite{b1} and

\begin{equation}
|\psi_{\nu=1}\rangle~=~\prod_{i,j}~(u_i v_j-v_i u_j)
\end{equation}
is the spin polarized ground state at $\nu=1$.

 The generalized
state for $\nu=1/m$, $m$ being an odd integer, is written as
\cite{rev11,h1}
\begin{equation}
|\psi_{\nu}\rangle~=~\prod_{i<j}~(u_i v_j-v_i u_j)^{1/{\nu}}
\end{equation}

However it should be added that at $\nu=1$ we have a
non-interacting spectrum which is never a correspondence for
$\nu=1/m$, with $m$ being an odd integer. Indeed FQH states
correspond to interacting systems.

The  skyrmion state can  be written as
\begin{equation}
|\psi\rangle~=~C^\prime \prod_k \left(\begin{array}{c}
  v_k \\
  -\alpha u_k
\end{array}\right)|\psi_{\nu}\rangle
\label{3}
\end{equation}

where the spin texture is included within the components $v_k$ and
$u_k$ and $0\leq\alpha\leq 1$.

Actually, if a smooth and monotonical function $g(\theta)$ is
defined with $g(0)=0$ and $g(\pi)=\pi$ then the skyrmion state can
be written as \cite{xh}
\begin{equation}
\hat{\phi}(\Omega)~=~\cos(g(\theta)-\theta)~\hat{e_r}+~\sin(g(\theta)-\theta)~\hat{e_\theta}
\label{eq17}
\end{equation}
where $\hat{e_r}$ and $\hat{e_\theta}$ are the basis vectors. The
size of a skyrmion is determined by the function $g(\theta)$ and
$g(\theta)=\theta$ describes the hedgehog skyrmion with spin in
the radial direction $\hat{r}$.

 This is the
skyrmion state with the constraint $|\hat{\phi}(\Omega)|=1$.

The quantum state for the classical skyrmion $\hat{\phi}(\Omega)$
can be written as
\begin{equation}
|\psi\rangle=~C.P\prod_k\left(\begin{array}{c}
  \sin{g(\theta_k)/2}~~e^{-i\phi_k/2} \\
  -\cos{g(\theta_k)/2}~~e^{i\phi_k/2}
\end{array}\right)|\psi_{\nu}\rangle
\label{5}
\end{equation}
where $C$ is the normalization constant, $P$ is the projection
operator to the lowest Landau level and  $g(\theta)$ controls the
size of the skyrmion \cite{xh}. From eqns. (\ref{3}) and (\ref{5})
it is seen that  $0\leq\alpha\leq 1$ is determined from
$g(\theta)$ and it controls the size of the skyrmion.

 The size of the skyrmion can be
defined by the position  where spin direction is perpendicular to
the radial (and spin) direction at either pole. With this
convention, the skyrmion size is
\begin{equation}
\theta_0= 2 arc\tan\alpha
\end{equation}
 which equals $\pi/2$ for the hedgehog
skyrmion with $\alpha=1$

Bychkov et. al. have shown \cite{by} that the quantum Hall
skyrmions consist of a core whose size is defined by the interplay
between the Zeeman and Coulomb energies and an additional length
scale $l_{sk}$ which determines the tail of the spin distribution.
This characteristic length is given by
\begin{equation}
  {l_{sk}}^{-2}=2\sqrt{2/{\pi}}|\tilde{g}|(a_B/l^3)
\end{equation}
where $\tilde{g}$ is the effective $g$ factor,
$a_B=\epsilon\hbar^2/me^2$ is the Bohr radius, $\epsilon$ being
the dielectric constant and $l$ is the magnetic length. It is
noted that as $\tilde{g}\rightarrow 0$, $l_{sk}\rightarrow
\infty$, we can take
\begin{equation}
  \cot{\theta/2}=x/x_0
\end{equation}
where $x$ and $x_0$ are two dimensionless parameters given by
$x=r/l_{sk}$ and $R=x_0
 l_{sk}$, $R$ being the size of the
skyrmion core region.

We can associate $\alpha$ with $x_0/x$ where $x_0\ll 1$ and $x\ll
1$ i.e. for $r$ far away from the exponential tail of the
skyrmion. It is pointed out that in the region $x_0\ll x\ll 1$,
$r$ is far outside the skyrmion core.

In the framework of $O(4)$ nonlinear $\sigma$-model the size of
the skyrmion is determined by the quartic stability term, known as
the Skyrme term. Indeed, taking the spin variable
${\bf{z}}=U{\bf{z_0}}$ where ${\bf{z}}_0=\left(\begin{array}{c}
  1 \\
  0
\end{array}\right)$
and $U\in SU(2)$, we may write the nonlinear sigma model
Lagrangian in terms of the $SU(2)$ matrices $U$ as \cite{sk}
\begin{equation}
L=~-\frac{M^2}{16}~(\partial_{\mu} U^{\dag}\partial_\nu
U)-\frac{1}{32 \eta^2}~[\partial_\mu UU^{\dag},\partial_\nu
UU^{\dag}]^2 \label{28}
\end{equation}
where $M$ is a constant of dimension of mass and $\eta$ is a
dimensionless coupling parameter. The $\alpha$ dependence may be
incorporated through $M$ and $\eta$ where these parameters are
taken as functions of $\alpha$.

 To have a geometrical interpretation of the Skyrme  term (eqn.(\ref{28})), we
note that it effectively corresponds to the vorticity of the
system which prevents it from shrinking it to  zero size. In fact
in an axisymmetric system where the anisotropy is introduced along
a particular direction through the introduction of a magnetic
monopole at the centre, the components of the linear momentum
satisfy a commutation relation of the form

\begin{equation}
[p_i,p_j]=i \mu ~\epsilon_{ijk}~\frac{x^k}{r^3}
\end{equation}

When the position space is a 3-sphere $S^3$ with a monopole  at
the centre, we can have a commutation relation of the form
\begin{equation}
[p_i,p_j]=\frac{i}{R}~\epsilon_{ijk}~p_k
\end{equation}
 where R is
proportional to the radius of the $S^3$. For a distorted sphere we
can consider $R$ as a functional form $R(\theta ,\phi)$
corresponding to the core radius of the skyrmion. We can define
the core size of the skyrmion such that $R=R_0(1-\a)$ where $R_0$
is the size of the  skyrmion having minimal energy.

In view of this, in $3+1$ dimensions we can generalize the
Lagrangian (\ref{11}) incorporating the Skyrme term in the form

\begin{equation}
\begin{array}{lcl}
L~
&=&\displaystyle{~2\pi\tilde{{\cal{J}}}^S_\mu~({\cal{A}}_\mu+{\cal{A}}^{[0]}_\mu)-~\frac{M^2}{16}~Tr(\partial_{\mu}
U^{\dag}\partial_\nu U)-\frac{1}{32 \eta^2}~Tr[\partial_\mu
UU^{\dag},\partial_\nu UU^{\dag}]^2}\\
&&\displaystyle{-\frac{\theta}{16\pi^2}~
{^{*}{\mathcal{F}}_{\mu\nu}}{\mathcal{F}}_{\mu\nu}-\frac{1}{4}{\mathcal{F}}_{\mu\nu}{\mathcal{F}}^{\mu\nu}}\\
\end{array}
\label{31}
\end{equation}

Here $\theta=g/c^2$ with $g=\nu e^2/h$ as Hall conductivity and
$~^{*}{\mathcal{F}}_{\mu \nu}$   is a Hodge dual given by
\begin{equation}
^{*}{\mathcal{F}}_{\mu\nu}
=\frac{1}{2}~\epsilon_{\mu\nu\lambda\sigma}{\mathcal{F}}_{\lambda\sigma}
\end{equation}

This is associated with $O(4)$ nonlinear sigma model where the
topological current is defined as

\begin{equation}
\tilde{{\mathcal{J}}}^S_{\mu}~=~\f{1}{24 \pi^2}~ \eps_{\m\n\a\b}
Tr(U^{-1}\pad_{\n}U)(U^{-1}\pad_{\a}U) (U^{-1}\pad_{\b}U)
\end{equation}
and ${\cal{A}}_{\mu}[{\cal{A}}^{[0]}_{\mu}]$ is a four vector.
 It is to be mentioned that we have taken the kinematic term
(second term in equation (\ref{31})) which takes into  account
spin waves, the Goldstone modes associated with a ferromagnetic
ground state.  The $SU(2)$ matrix $U$ is here defined as
\begin{equation}
U=n_0 I + \bf{n}.\overrightarrow{\tau}
\end{equation}
where the chiral fields $n_0$,$n_1$,$n_2$ and $n_3$ satisfy the
relation $\sum n_i^2=1$.

It may be noted that in $2+1$ dimension, we have the normalized
$3$-vector field ${\bf n}$ with $\sum n^2_i=1$ where $n_i$
corresponds to the local spin direction. In the $O(4)$ model $n_i
(i=1,2,3)$ corresponds to this spin direction which live on the
$2$-dimensional surface of the sphere where the extra field $n_0$
helps us to consider three "boost"  generators in $(n_0, n_i)$
planes. In view of this, we can consider two types of generators
such that the generator $M_k$ rotates the $3$-vector $\bf{n}(x)$
to any chosen axis and the boost generators $N_k$ would mix $n_0$
with the components of $\bf{n}$. We can now construct the
following algebra

$$[M_i,M_j]= {i \epsilon_{ijk} M_k}$$
$$[M_i,N_j]= {i \epsilon_{ijk} N_k}$$
\begin{equation}
[N_i,N_j]= {i \epsilon_{ijk} M_k}
\end{equation}
which is locally isomorphic  to the Lie algebra of the $O(4)$
group. This helps us to introduce the left and right generators
$$L_i=\frac{1}{2}(M_i-N_i)$$
\begin{equation}
R_i=\frac{1}{2}(M_i+N_i)
\end{equation}
which satisfy
$$[L_i,L_j]= {i \epsilon_{ijk} L_k}$$
$$[R_i,R_j]= {i \epsilon_{ijk} R_k}$$
\begin{equation}
[L_i,R_j]=0
\end{equation}

Thus the algebra has split into two independent subalgebras each
isomorphic to a $SU(2)$ algebra and corresponds to the chiral
group $SU(2)_L \otimes SU(2)_R$. The left and right chiral group
can now be taken to be associated with two mutually opposite
orientations of the magnetization vector which resides on the $2$D
surface of the sphere.

 It may be mentioned that the $P$ and $T$ violating $\theta$-term
$(-\frac{\theta}{16\pi^2}~
{^{*}{\mathcal{F}}_{\mu\nu}}{\mathcal{F}}_{\mu\nu})$ in $3+1$
dimensions corresponds to the Chern-Simons term in $2+1$
dimension. This term is related to chiral anomaly and we have the
Pontryagin index given by \cite{db}
\begin{equation}
 q~=~2\mu=~-{\f{1}{16 \pi^2}}~\int Tr ^{*} {\mathcal{F}}_{\m\n} {\mathcal{F}}_{\m\n} d^4 x
\label{38}
\end{equation}

It may noted that in $2+1$ dimensions,  the topological current
\begin{equation}
J_\mu=\frac{1}{8\pi}~\epsilon_{\mu\nu\lambda}{\bf{n}}.(\partial_\nu
{\bf{n}}\times
\partial_\lambda {\bf{n}})
\end{equation}
may be related to the gauge field current
\begin{equation}
  J_\mu=\epsilon_{\mu\nu\lambda} \nabla_\nu {\cal{A}}_\lambda
\end{equation}
which helps us to write the Chern-Simons term as
\begin{equation}
W=\frac{\theta}{4\pi}\int d^3 x
\epsilon_{\mu\nu\lambda}{\cal{A}}_\mu \nabla_\nu {\cal{A}}_\lambda
\end{equation}

Evidently this relates to the Hopf invariant which can be
explicitly written in terms of $SU(2)$ matrix $U(x)$ which rotates
the vector ${\bf n}$ to the chosen (third) axis ${\bf n}.{\vec
\tau}=U^{-1} \tau_3 U$ and the Hopf invariant is given by
\begin{equation}
H=\frac{\epsilon_{\mu\nu\lambda}}{24 \pi^2}\int d^3 x
Tr(U^{-1}\partial_{\mu}U)(U^{-1}\partial_{\nu}U)
(U^{-1}\partial_{\lambda}U)
\end{equation}
which is a degree of mapping of the (2+1) dimensional space time
into $SU(2)$ and is given by the homotopy $\pi_3(S^3)=Z$. In $3+1$
dimensions, if we consider Euclidean four dimensional space-time
such that ${\mathcal{F}}_{\mu \nu}=0$ at all points on the
boundary $S^3$ of a certain volume $V^4$ inside which
${\mathcal{F}}_{\mu\nu}\neq 0$ then the gauge potential tends to
be a pure gauge in the limiting case towards the boundary i.e. we
have \beq {\cal{A}}_\mu = U^{-1}\partial_\mu U,~~~~~~~~~~~~~~U\in
SU(2)\eeq

This then helps us to write the topological charge of quantum Hall
skyrmions given by eqn.(38) as the winding number
 \beq Z~=~ {\f{1 }{24
\pi^2}}~\int_{S^3} dS_{\m}~ \eps^{\m\n\lam\s} [(U^{-1} \pad_{\n}
U)(U^{-1} \pad_{\lam} U)(U^{-1} \pad_{\s} U)] \eeq which is given
by the homotopy $\pi_3(S^3)=Z$ and the electric charge is given by
$\nu e Z$. We observe here that the Pontryagin  index given by
$q=2\mu$ in eqn.(\ref{38}) introduces the Berry phase for quantum
Hall states as $\mu$ represents magnetic monopole strength.
Indeed, the flux through the sphere is $2 \mu$ and the phase is
given by $e^{i\phi_B}$ where $\phi_B=2 \pi \nu$ (number of flux
quanta enclosed by the loop). When a $Z=1$ skyrmion  is moved
around a closed loop it acquires a Berry phase $2\pi\nu N$ where
$N$ is the number of skyrmions enclosed by the loop. To find the
spin and statistics, we consider a process which exchanges two
skyrmions in the rest frame of one of the skyrmions. This exchange
effectively corresponds to the other skyrmion moving around the
first in a half circle and hence it picks up a phase $\pi~\nu$
which is the statistical phase of a skyrmion. For $\nu=1$ it is a
fermion and $\nu=1/m$, $m$ being an odd integer it corresponds to
an anyon in planar geometry. In general, the spin of the skyrmion
having charge $\nu e Z$ is given by $\nu Z/2$.

\section{Skyrmion Size and Energy}

From our above discussions, we now observe that the energy of a
free skyrmion depends on the following two terms in the Lagrangian
(\ref{31})
\begin{equation}
L=-~\frac{M^2}{16}~(\partial_{\mu} U^{\dag}\partial_\nu
U)-\frac{1}{32 \eta^2}~[\partial_\mu UU^{\dag},\partial_\nu
UU^{\dag}]^2 \label{45}
\end{equation}

The static nonlinear sigma model Lagrangian corresponding to eqn.
(\ref{45}) gives rise to the energy integral as
\begin{equation}
E=\int d^3 x \{\frac{M^2}{16}Tr (\nabla U^{\dag}\nabla U) +
\frac{1}{32 \eta^2}Tr [\partial_i U U^{\dag},\partial_j U
U^{\dag}]^2\}
\end{equation}
where $(i,j=1,2,3)$

 To compute the energy, we take the Skyrme
ansatz \cite{sk}
\begin{equation}
U(x)=exp~[i F(r)\overrightarrow{\tau}.{\bf{\hat{x}}}]
\end{equation}
where $\overrightarrow{\tau}$ are Pauli matrices,
$\bf{\hat{x}}=\frac{x_i}{r}$ and $F(0)=\pi$ and $F(r) \rightarrow
0$ as $r\rightarrow \infty$; $r$ is the spherical radius in
$3$-dimensional manifold. We explicitly write $U=\cos{F(r)}+i
\overrightarrow{\tau}.{\bf{\hat{x}}} \sin{F(r)}$ with
$\cos{F(r)}=\frac{1-(r/R)^2}{1+(r/R)^2}$,~~
$\sin{F(r)}=\frac{2(r/R)}{1+(r/R)^2}$. The energy integral becomes
\begin{equation}
E(R)=4\pi^2 M^2 R~ I_1+~2 \pi^2\frac{I_2}{\eta^2 R}
\end{equation}
where $$I_1 =\frac{1}{\pi}\int^{\infty}_0
dx[\sin^2{F(r)}+x^2~(\frac{\partial F}{\partial x})^2]=~3.0$$ and
$$I_2 =\frac{1}{\pi}\int^{\infty}_0
dx[\frac{\sin^4{F(r)}}{x^2}+~\sin^2{F(r)}(\frac{\partial
F}{\partial x})^2]=~ 1.5$$ where $x=r/R$. This gives the
expression of energy
\begin{equation}
E(R)=12 \pi^2 M^2 R + \frac{3 \pi^2}{\eta^2 R} \label{49}
\end{equation}

The minimum of energy $E(R)$ is found from the relation
\begin{equation}
  \frac{\partial E(R)}{\partial R}=12 \pi^2 M^2-\frac{3 \pi^2}{\eta^2
  R^2}=0
\end{equation}
which gives for $E_{min}$, the size as
\begin{equation}
  R_0=\frac{1}{2M \eta}
\label{51}
\end{equation}
and the energy
\begin{equation}
E_{min}=E(R_0)=12 \pi^2 M/ \eta \label{52}
\end{equation}

It is to be noted that the coupling parameters $M$ and $\eta$ are
functions of $\a$ such that in the limit $\a \rightarrow 0$,
$M(\a) \rightarrow 0$ and $\eta(\a) \rightarrow 0$ but $M/ \eta$
is fixed.

We may compare $E_{min}$ with the scale invariant energy
\begin{equation}
E(0)=\frac{\sqrt{\pi}}{4\sqrt{2}}\frac{e^2}{\epsilon l} \label{53}
\end{equation}
which is obtained from eqn. (15) with $g=0, V=0$, in the
Lagrangian (\ref{13}). In this case energy is given by the {\it
pure nonlinear sigma model}~: the size of the skyrmion is infinite
and the energy is independent of the size. So we may write
\begin{equation}
E(R_0)=12 \pi^2 M/ \eta= \frac{\sqrt{\pi}}{4\sqrt{2}}~e^2/\epsilon
l
\end{equation}

Away from $g=0$, skyrmions acquire a size. Now matching this value
for $g \ll 1$ so that the form of the solution in the core region
is determined by the scale invariant term alone, we take
$R=R_0(1-\a)$  and we have
\begin{equation}
\begin{array}{lcl}
E(R) &=&\displaystyle{\frac{6 \pi^2 M}{\eta}
[(1-\a)+\frac{1}{(1-\a)}]}\\
&&\displaystyle{=\frac{\sqrt{\pi}}{4\sqrt{2}}~e^2/\epsilon l~
\frac{1}{2}~[2+\frac{\a^2}{(1-\a)}]}\\
\end{array}
\label{55}
\end{equation}

 We can compare this with the standard result for energy with
 $Z=1$, \cite{s3}
\begin{equation}
E(g)=\frac{\sqrt{\pi}}{4\sqrt{2}}\frac{e^2}{\epsilon l}
[1+\frac{24.9}{4 \pi}(\tilde{g}|ln\tilde{g}|)^{1/3}]
\end{equation}
where $\tilde{g}=(g \mu_B B)/(e^2/\epsilon l)$ so that we can
equate
\begin{equation}
\frac{1}{2}[2+\frac{\a^2}{(1-\a})]=1+\frac{24.9}{4
\pi}(\tilde{g}|ln\tilde{g}|)^{1/3}
\end{equation}
which suggests
\begin{equation}
\frac{\a^2}{1-\a}=\frac{24.9}{2 \pi}(\tilde{g}|ln\tilde{g}|)^{1/3}
\label{58}
\end{equation}

\begin{figure}
  \centerline{\epsfxsize=4.0in \epsfbox{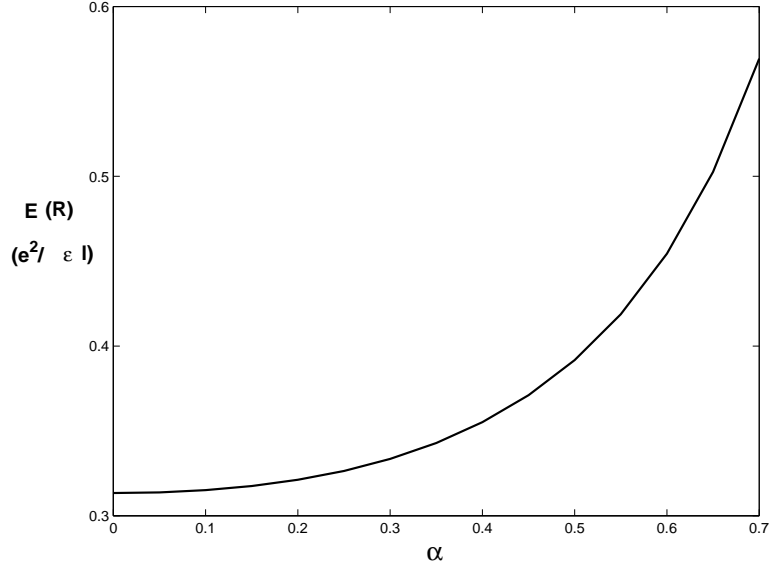}}\label{f1}
  \caption{The dependence of skyrmion energy on size with respect to the variation of the parameter $\a$ (eqn.(\ref{55})) }
\end{figure}

\begin{figure}
  \centerline{\epsfxsize=4.0in \epsfbox{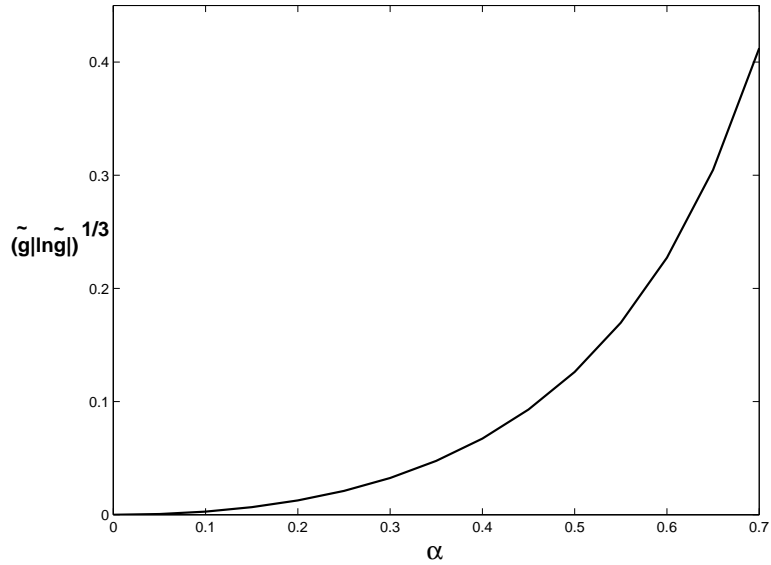}}\label{f2}
  \caption{The relationship of $(\tilde{g}|ln\tilde{g}|)^{1/3}$
  with $\a$ as given by the eqn.(\ref{58})}
  \end{figure}
We must mention here that  though the value of $\a$ is constrained
in the region $0 \leq \a\leq 1$ and $\a=1$ corresponds to the
hedgehog skyrmion with spin in the radial direction but  eqn.
(\ref{55}) suggests that $\a=1$ is a singularity point. Indeed,
the relation $R=R_0 (1-\a)$ gives a nonzero size for $\a=1$ when
$R_0$ is infinite. So in this case matching the minimum energy for
$g=0$, $V=0$ and away from $g=0$ with $g \ll 1$ will not be
meaningful. Indeed the solution $\cot{\theta/2}=x/x_0$ is valid
except for a very narrow region around $x=x_0$ \cite{by}. This
implies that a small region around $\a=1$ will not give meaningful
result for size and energy of the skyrmion. For comparison we have
computed the energies of skyrmions in terms of $\a$ and it is
found that up to $\a=0.7$ we have reasonable values of $\tilde{g}$
beyond which $\tilde{g}$ becomes large enough to be compatible
with quantum Hall skyrmions [Fig. 1].

This suggests that in spherical geometry within the framework of
$O(4)$ nonlinear sigma model in $3+1$ dimensional manifold we can
determine the size of the skyrmion incorporating the Skyrme term
in the Lagrangian. The ratio of the Zeeman energy and Coulomb
energy can be encoded in it [Fig. 2].

It may be  added here that very recently W\'{o}js and Quinn
\cite{qu} presented the numerical results for the spin excitation
spectra of integral and fractional Hall systems in spherical
geometry.

%\newpage

\section{Discussion}
Many authors have considered quantum Hall skyrmions in spherical
geometry and studied different aspects of the system in terms of
$O(3)$ nonlinear sigma model with success. In all these
approaches, the electrons and hence the 3 (constrained) order
parameter fields reside on the $2$-dimensional surface of the
$2+1$-dimensional $O(3)$ model. To study the static properties of
the skyrmions  we have used here the $3+1$-dimensional $O(4)$
nonlinear sigma model which actually builds on a different
manifold and possesses different number of order parameter fields.
The extra order parameter field helps us to consider two
independent $SU(2)$ algebras depicting two mutually opposite
orientations of the magnetization vector which lives on the
$2$-dimensional surface. In this framework with the homotopic
analysis we have shown the existence/non-existence of skyrmions in
fully polarized, unpolarized and partially polarized quantum Hall
states \cite{bdp}.
 The Skyrme term which gives
rise to the stability of the soliton  determines the size of
quantum Hall skyrmions. Indeed, the distortion of the spherical
shape can be incorporated through the angular dependence which can
be conveniently introduced through a parameter $\a$ with $0 \leq
\a \leq 1$. Excepting a small region around $\a=1$, we can compare
the result with the conventional $2D$ formalism when the size is
determined by the Zeeman energy and Coulomb energy for small $g$.
This helps us to encode the effect of the Zeeman energy and
Coulomb energy in the parameter $\a$ and the size can be
determined from the relation $R=R_0(1-\a)$ where $R_0$ corresponds
to the size which is characteristic of the minimum energy
$E(R_0)$. In $2D$ formalism this minimum energy corresponds to
$\tilde{g} \ll 1$ which may be matched with the scale invariant
value when the energy is contributed by the spin stiffness term
for $g=0, V=0$.

We have also shown that the spin and statistics of the skyrmion is
given by the $\theta$-term. The skyrmion charge is given by $\nu e
Z$ where $Z$ is the winding number associated with the homotopy
$\pi_3(S^3)=Z$ and the spin is given by $\nu Z/2$. For $Z=1$, the
skyrmion, when moving around a closed loop picks up a phase $2\pi
\nu N$ where $N$ is the number of skyrmions enclosed by the loop.
In a process which exchanges two skyrmions in the rest frame of
one of the skyrmions, the exchange corresponds to the other
skyrmion moving about the first in a half circle and hence it
picks up a phase $\pi \nu$ which is the statistical phase.

\newpage

%{\bf {References}}

\ed